\documentstyle[twocolumn,epsfig,aps]{revtex}

\begin{document}

\draft
\title{Statistics of stable marriages}
\author{Michael Dzierzawa
\footnote{Corresponding author\\
e-mail: Michael.Dzierzawa@Physik.Uni-Augsburg.de
}
}
\address{Institute of Physics, University of Augsburg,
 D-86135 Augsburg, Germany}
\author{Marie-Jos\'e Om\'ero}
\address{Computer Science Department, University of Geneva,
CH-1211 Gen\`eve 4, Switzerland}
\date{\today}
\maketitle

\begin{abstract}
In the stable marriage problem $N$ men and $N$ women have to be matched
by pairs under the constraint that the resulting matching is stable.
We study the statistical properties of stable matchings in the large $N$
limit using both numerical and analytical methods. Generalizations of
the model including singles and unequal numbers of men and women
are also investigated.
\end{abstract}
\pacs{PACS: 05.20.-y, 01.75.+m, 02.50.Le\\
{\sl Keywords}: Stable marriage, optimization }

\section{Introduction}

The study of complex systems that are built up of many individuals 
interacting according to relatively simple laws
has recently attracted a lot of interest among physicists.
Concepts and methods developed in statistical physics
have successfully been applied to describe collective and
nonlinear phenomena arising in such systems. Studies on
traffic flow, stock markets, voting, flocking birds, evolution
etc. are only a few examples to mention here.

In the stable marriage problem the elements of two sets have to be matched
by pairs. As possible applications one could think of job seekers and employers, 
lodgers and landlords, or simply men and women who want to get married.
For the sake of clearness and simplicity
we will exclusively refer to the paradigm of marriage 
in this paper.
The purpose of most studies of the stable marriage problem has so far been 
to unravel the underlying mathematical structure and
to develop new or improve existing algorithms for finding stable marriages
\cite{Knu97,Gus89}.
Only recently it has been recognized that the stable marriage problem
has many features in common with classical disordered systems in 
statistical physics
like spin glasses, in particular the element of frustration and
the existence of many competing states \cite{Ome97}. 
Applying methods developed in statistical mechanics,
Nieuwenhuizen \cite{Nie98} solved several variants of the stable
marriage problem, including bachelors and polygamy.
In his approach, Nieuwenhuizen focussed on the properties of
globally optimal matchings which are advantageous for the society as 
a whole, but not necessarily for all
individuals. Here we consider the opposite limit 
where all individuals pursue only their own egoistic objectives
regardless of the consequences for others. 
Under these conditions, only matchings that are
stable with respect to the actions of all individuals are relevant.
In the framework of game theory this kind of stability
is termed Nash equilibrium.

Our paper is organized as follows. In Section 2 we introduce the stable
marriage model and set up the notation. The Gale-Shapley algorithm and
the properties of the stable matchings obtained with it are presented in 
Section 3. In Section 4 we analyze the set of all stable matchings.
Generalizations of the model, introducing an acceptance threshold and unequal
number of men and women are discussed in Sections 5 and 6, respectively.
Finally, we summarize our results in Section 7.

\section{The stable marriage model}

We consider a society consisting of
$N$ men and $N$ women who have to be matched by pairs.
The cost for man $i$ to be married with woman $j$ is $x(i,j)$ and
the cost for woman $j$ to be married with man $i$ is $y(j,i)$.
In order to investigate statistical properties of the model we assume that all
cost functions are independent random variables uniformly distributed between
$0$ and $1$ \footnote{In Ref. \cite{Nie98} a different probability distribution
has been considered, namely $\rho(x) \propto x^r\, \exp(-x)$.
The case $r=0$ corresponds essentially to our choice.}.
Since all individuals simultaneously attempt to optimize their
own benefit by being married with a partner of lowest possible cost, conflicting
wishes arise, and in general it is impossible to find a matching where everybody
is perfectly satisfied. 
Let us assume that a matching is formed where man $i$ is married with 
woman $j$. We define $x_i \equiv x(i,j)$ and $y_j \equiv y(j,i)$,
i.e. $x_i$ is the cost for man $i$ in this particular matching and
$y_j$ the corresponding cost for woman $j$.
Instead of {\em cost} we will frequently use the expression {\em energy} 
in order to emphasize the analogy with physical systems where 
the energy is the quantity to be minimized in the groundstate.
For later convenience, we introduce the 
total energies of men and women by
\begin{equation}
X = \sum_{i=1}^N x_i  ~~~~~~~~ Y = \sum_{j=1}^N y_j
\end{equation}
In search of the rules according to which marriages can be formed
or broken, essentially two points of view have emerged:
In the first one, the goal is to find the globally best matching,
i.e. the one where the total energy $U = X + Y$ is minimal. This problem has
recently been studied by Nieuwenhuizen \cite{Nie98} using techniques 
borrowed from statistical physics of disordered systems,
such as the replica trick.
The second point of view, that we will adopt, emphasizes the role of individuals
as decision-makers. As long as there is no supervising body like church or 
government that dictates who has to be married with whom
no individual is obliged
to stay in an unsatisfactory relationship if he or she can find a better one.
This leads us to the notion of stability.
A matching is called {\em stable} if there is no man and no woman who 
prefer each other to their actual partners.
In other words, a matching is {\em unstable}, if there exists
a man $i$ and a woman $j$, who are {\em not} married with each other, such that
\begin{equation}
x_i > x(i,j) ~~  {\rm and} ~~ y_j > y(j,i)
\end{equation}
The particular significance of stable matchings lies in the fact that they cannot
be broken, neither by the action of a single pair nor by 
any coalition of men and women.
In contrast to the globally best matching which can only be found with
the help of computers for very small numbers of men and women, there
exist powerful algorithms to search for stable matchings, even
for systems of several thousands of individuals.
The simplest of these algorithms, introduced by Gale and Shapley
some time ago \cite{Gal62}, will be presented in the next section.

\section{The Gale-Shapley algorithm}

In the {\em man-oriented} Gale-Shapley (GS) algorithm only men make proposals 
while women accept or reject. In the beginning every man makes up
a list where he places the women
according to his preferences. The smaller the cost $x(i,j)$ 
the higher the position of woman $j$ on the list of man $i$.
Then a series of proposals starts. Whenever a man is not married
he makes a proposal to the woman with the highest rank among those to
whom he has not yet proposed.
If the woman is not engaged she accepts the proposal and
they get married. If the woman is already married, she accepts
if she prefers the suitor to her husband. 
In this case she divorces and gets married with the proposer.
The algorithm stops when the last woman has received an offer and gets married.
The number of proposals $r$ that a man has to make on the average 
during the execution of the GS algorithm
has been calculated in \cite{Ome97}. The result is
\begin{equation}
r \simeq \log N + {\rm C} 
\label{NPROP}
\end{equation}
where ${\rm C} = 0.5772...$ is Euler's constant and corrections are
negative and of order $(\log N)^2 / N$.
Amazingly, the knowledge of $r$ is already
sufficient to calculate the total energy of men ($X$)
and women ($Y$) in the GS matching.
With every proposal the energy of one man increases by $\simeq 1/N$ since
he has to move forward one step on his preference list. 
Consequently, the total energy of men is on the average the same as the
number of proposals that every man makes, i.e. $X = r$.

Each woman also receives on the average $r$ proposals.
Keeping only the best offer i.e. the smallest out of $r$ 
random numbers between
$0$ and $1$ yields a best value of $y = 1 / (r + 1)$.
Taking into account that the number of proposals that a woman receives is 
not fixed, but distributed according to a binomial distribution
yields $y = (1 - \exp(-r))/ r \approx 1/r$ (see Appendix B). 
Using Eq. (\ref{NPROP}) we obtain the following estimates for the 
energies of men and women in the man-oriented GS matching
\begin{equation} 
X \simeq \log N + {\rm C} 
~~~~~~~ Y \simeq \frac{N}{\log N + {\rm C}}
\label{XY}
\end{equation}
which implies the relation $X\, Y = N$.

\begin{figure}
\centerline{\epsfig{file=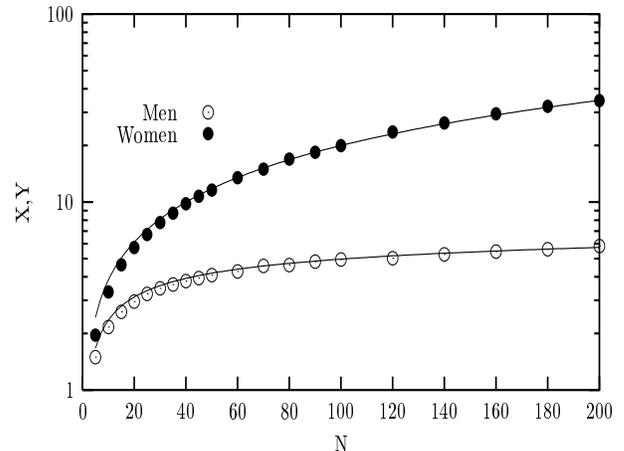,width=8cm,height=6cm,angle=0}}
\caption{Average energy of men and women in the man-oriented
Gale-Shapley algorithm as a function of $N$.}
\label{gs}
\end{figure}

Fig. \ref{gs} shows the average energies $X$ and $Y$ of men and women obtained with
the man-oriented GS algorithm as function of $N$.
Each data point represents an average 
over several thousand realizations. A logarithmic scale is used for the 
energy axis in order to highlight the good
agreement with the analytical result of Eq.(\ref{XY})
even for small values of $N$.

Of course, everything that has been said about the man-oriented
GS algorithm is also valid in the woman-oriented
version where the roles of men and women are interchanged, i.e. women
propose and men judge.
It has been shown \cite{Gus89} that
the stable matching obtained from the man--oriented 
execution of the GS algorithm 
is man--optimal in the sense that no man can have lower energy in any other 
stable matching. Similarly, it is the worst possible stable matching 
for women. Therefore the energies obtained by performing the GS
algorithm provide an upper and a lower bound for the possible 
energy range of {\em all} stable matchings.
Furthermore, if the man and the woman--optimal matching happen
to be identical we know for sure that in this case there exists only one 
single stable matching.

\section{The set of all stable matchings}

If $N$ is large, there are generally many other stable matchings besides
the man and the woman-optimal matchings that can be obtained with the
GS algorithm. Although the {\em maximum} number 
of stable matchings in a system of size $N$ is not known, 
there exist lower bounds \cite{Gus89}; e.g. for $N = 32$ the
maximum number of stable matchings is larger than $10^{11}$ (!).

In order to calculate the 
{\em average} number of stable matchings we consider an arbitrary matching
of $N$ men and $N$ women. This matching is
unstable, if there exists a man $i$ and a woman $j$ who are not married with each
other, and whose mutual costs $x(i,j)$ and $y(j,i)$ are both
lower than the ones in the existing matching, $x_i$ and $y_j$, respectively.
The probability that man $i$ and woman $j$ prefer to
stay with their respective partners is therefore $p_{ij} = 1 - x_i\,y_j$ and
the whole matching is stable with probability
\begin{equation}
P = \int_0^1 {\rm d}^Nx
\int_0^1 {\rm d}^Ny \,\prod_{i\neq j} (1 - x_i\,y_j)
\label{KNUTH}
\end{equation}
Although this $2N$-fold integral, originally derived by Knuth \cite{Knu97},
looks quite simple, it cannot be done exactly unless $N$ is very small.
Using rather involved probabilistic arguments, Pittel \cite{Pit89} has derived
an asymptotic formula,
\begin{equation}
P \simeq \frac{\log N}{{\rm e} \Gamma(N)} 
\label{PITTEL}
\end{equation} 
valid in the limit $N \rightarrow \infty$. However, comparing Eq.
(\ref{PITTEL}) with numerical simulations we found rather serious 
discrepancies (see Fig. 2).
In order to remove these discrepancies and to provide a more
transparent approach
we replace the factors $1 - x_i\,y_j$ in Eq. (\ref{KNUTH}) by $\exp(-x_i\,y_j)$,
which is justified since the dominant contribution to the integral
comes from the region where the products $x_i y_j$ are small. Dropping 
the constraint $i\neq j$ we arrive at the approximation
\begin{equation}
\tilde{P} = \int_0^1 {\rm d}^Nx
\int_0^1 {\rm d}^Ny \,\exp(-X\,Y)
\label{PS}
\end{equation}
where the variables $X = \sum_{i=1}^N x_i$ and $Y = \sum_{j=1}^N y_j$
are the energies of men and women, respectively. Eq. (\ref{PS}) can also
be viewed as the partition sum of two species of particles
confined to the interval $[0,1]$
whose interaction is the product of their center of mass coordinates.
It is convenient to replace the integral over the variables $x_i$ and $y_j$  
by an integral over $X$ and $Y$
\begin{equation}
\tilde{P} = \int_0^N {\rm d} X \int_0^N {\rm d} Y \, \rho(X) \rho(Y)
\,\exp(-X Y)
\label{tilde}
\end{equation}
where the probability distribution of $X$ is asymptotic to
\begin{equation}
\rho(X) \simeq \frac{X^{N-1}}{\Gamma(N)} \left(1 - \exp(-\frac{N}{X})\right)^N
\label{rho}
\end{equation}
for small $X$ (see Appendix A). 
Inserting Eq. (\ref{rho}) in Eq. (\ref{tilde})
and introducing $t = X Y$ as new integration variable we
obtain
\begin{eqnarray}
\nonumber
\tilde{P} &=& \Gamma(N)^{-2} \int_0^{\frac{N}{\log N}} \,
{\rm d} X \,X^{-1} \int_0^{\frac{N\, X}{\log N}} \, {\rm d} t \; t^{N-1}
 \, \exp(-t)\\
 \nonumber
 &\simeq & \Gamma(N)^{-1} \int_{\log N}^{\frac{N}{\log N}} \,
{\rm d} X \,X^{-1} \\
& = & \Gamma(N)^{-1} \left( \log N - 2 \log(\log N) \right)
\end{eqnarray}
where we have assumed that the factor $\left(1 - \exp(-\frac{N}{X})\right)^N$
imposes a cut-off $\simeq N / \log N$ on the $X$-integration and the same holds for $Y$.
The integral over $t$ yields $\Gamma(N)$, provided that
$X > \log N$. Comparison with Eq. (\ref{PITTEL}) shows
that the ratio $\tilde P / P$ approaches ${\rm e} = \exp(1)$ in the limit $N
\rightarrow \infty$ which we also checked numerically.
Since there are in total $N ! = \Gamma(N+1)$
matchings the average number of stable matchings is given by
\begin{equation}
S \simeq \frac{N}{\rm e} \left( \log N - 2 \log(\log N) \right)
\label{NSTAB}
\end{equation}

\begin{figure}
\centerline{\epsfig{file=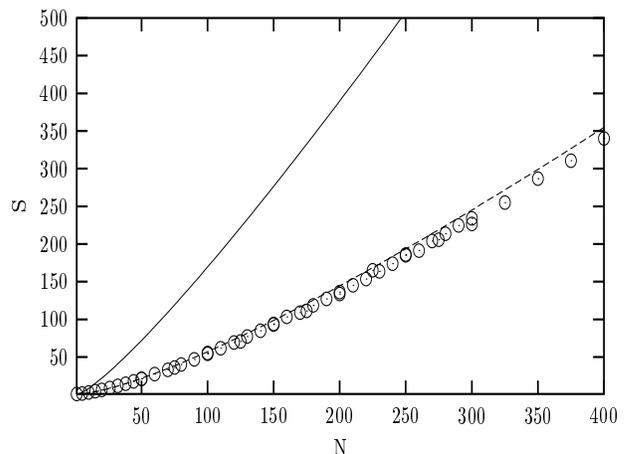,width=8cm,height=6cm,angle=0}}
\caption{Average number of stable matchings $S$ as function of $N$.
Numerical simulations (circles) in comparison with the analytic expression
of ref. [6] (solid curve) and our result of Eq.(\ref{NSTAB}) (dashed curve).}
\label{sofn}
\end{figure}

$S$ is plotted as function of $N$ in Fig. \ref{sofn}. 
The circles
are data obtained from numerical simulations using
the algorithm of Gusfield and Irving \cite{Gus89} that allows the determination
of all stable matchings. Each data point represents an average
over 200 random realizations of the preference lists. The dashed
curve is the result of Eq. (\ref{NSTAB}) and the solid curve is
the asymptotic formula derived by Pittel \cite{Pit89}. Note the
large error made by neglecting the $\log(\log N)$ term.

In the following we investigate how the energies of men and
women are correlated in stable matchings. 
From Eq. (\ref{tilde}) we see that 
$\rho(X,Y) = \rho(X) \rho(Y) \exp(-X Y)$ can be considered
as the probability density of finding a stable matching with energies
$X$ and $Y$. Due to the factor $\left(1 - \exp(-\frac{N}{X})\right)^N$
that appears in $\rho(X)$ the occurrence of stable matchings
with $X > N / \log N$ is very unlikely, and
by symmetry, the same cutoff exists for $Y$. 
This observation is in accordance with our analysis of the GS algorithm,
where we found that the energy of women in the man-optimal stable
matching (i.e. the worst possible for women) is 
$\simeq N / \log N$, and vice versa.
In the region $\log N < X, Y < N / \log N$ the probability distribution 
simplifies to $\rho(X,Y) \propto (XY)^{N-1} \exp(-X Y)$.
It depends only on the product of men's and women's 
energies and is sharply peaked around the curve $X Y = N$ for large values of $N$.
Mean-value and standard deviation of the product $X Y$ are given by
\begin{equation}
\langle XY \rangle = N, ~~~~ \sigma(XY) = \sqrt{N}
\end{equation}
where $\langle \ldots \rangle$ denotes the average calculated with $\rho(X,Y)$.
At the symmetric point $X = Y = \sqrt{N}$ the width of the distribution
is ${\rm O}(1)$ which means that relative fluctuations go to zero for
$N \rightarrow \infty$.


\begin{figure}
\centerline{\epsfig{file=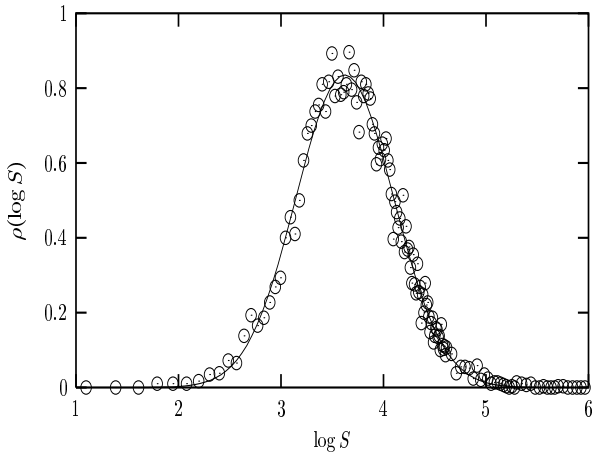,width=8cm,height=6cm,angle=0}}
\caption{Probability distribution of $\log S$ for $N = 50$. Numerical simulations
(circles) fitted by a Gaussian.}
\label{logs}
\end{figure}

\begin{figure}
\centerline{\epsfig{file=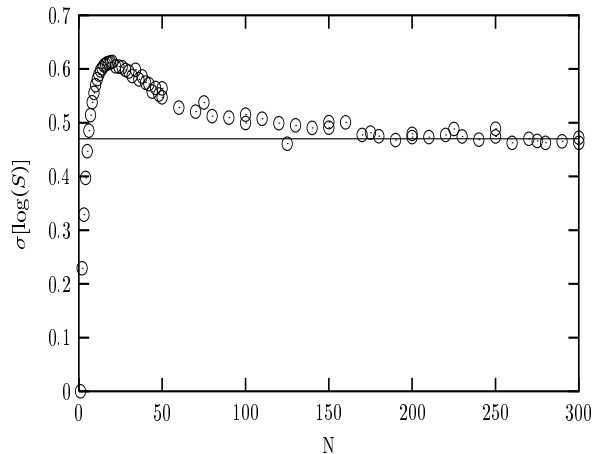,width=8cm,height=6cm,angle=0}}
\caption{Standard deviation $\sigma$ of $\log S$ as function of $N$. The numerical results 
(circles) seem to converge to the solid line $\sigma_{\infty} \approx 0.48$ 
for $N \rightarrow \infty$.}
\label{width}
\end{figure}

Finally, we investigate the probability distribution
of the number of stable matchings using numerical simulations.
While $S$ has a very asymmetric distribution which decays quite slowly 
for large values of $S$, the distribution of $\log S$ can be nicely
fitted with a Gaussian, as shown in Fig. \ref{logs} for $N = 50$.
Remarkably, the width of the 
distribution, i.e. the standard deviation $\sigma(\log S)$
approaches a constant value $\sigma_{\infty} \approx 0.48$ for large $N$ 
as shown in Fig. \ref{width}.
We have no argument to explain this universal behavior.

\section{Acceptance threshold}

Up to now we have assumed that every person {\em has} to get married.
In the following we soften this condition by introducing an acceptance
threshold $\Delta < 1$ which is the energy of individuals who remain single.
In principle, $\Delta$ could be chosen at random for all individuals,
or at least be different for men and women, but
for the sake of simplicity we consider only the case where $\Delta$ is
the same for all men and all women.

Let us consider a situation where $\Delta$ is sufficiently large
that singles do not occur. 
The GS algorithm is readily generalized to the case with acceptance threshold. 
In the man-oriented version the only difference is that now women
-- even when unmarried --
refuse proposals from men whose energy is higher than $\Delta$.
This increases the average number of proposals that are necessary to
arrive at a stable matching by a factor of $1/\Delta$.
The generalization of Eq. (\ref{XY}) to the case $\Delta \neq 1$ reads
\begin{equation}
X(\Delta)  \simeq  \frac{\log N + {\rm C}}{\Delta} ~~~~~~~~~
Y(\Delta)  \simeq  \Delta \, \frac{N}{\log N + {\rm C}}
\label{XYDEL}
\end{equation}
Both energies coincide for
\begin{equation}
\Delta_{\rm c} = \frac{\log N + C}{\sqrt{N}}
\end{equation}
and we conclude that for $\Delta < \Delta_c$ there exist only few or one 
single stable matching. This fact can be used to generate approximately
sex-fair stable matchings -- which is difficult to achieve
by other means --
using the GS algorithm and setting $\Delta = \Delta_c$.
 
In the limit when $\Delta$ is sufficiently small it becomes
favorable for a certain fraction of men and women 
to remain single. In this case it is hard to calculate the average
number of proposals $r$ explicitly 
since the series of proposals is not only terminated when all men are 
married but also if no acceptable women are available. On the other hand
the energies of men and women coincide in this regime since there exists
only one stable matching.
Let us assume that
a woman has received $r$ proposals out of which she keeps only the best one.
If none of the proposals are better than the threshold $\Delta$ she
remains single at the cost of $\Delta$. Analyzing the probability distribution
of the best offer (see Appendix B) yields the average energy of women
\begin{equation}
Y = \frac{N}{X} \left(1 - \exp(-X \Delta) \right)
\end{equation}
From the condition $X = Y$ we obtain
\begin{equation}
X^2 = N \left(1 - \exp(-X \Delta) \right)
\label{XSC}
\end{equation}
In general this equation can only be solved numerically.
The limiting cases are 
\begin{equation}
X(\Delta) \simeq \left\{ 
\begin{array}{lll}
N \Delta  & {\rm ~~~for~} & \Delta \ll \frac{1}{\sqrt{N}}\\
\sqrt{N}  & {\rm ~~~for~} & \Delta \gg \frac{1}{\sqrt{N}} 
\end{array}
\right.
\end{equation}
which means that for sufficiently small $\Delta$ it is advantageous for almost
everybody to remain single
(at the energy $\Delta$) while for $\Delta \gtrsim 1/\sqrt{N}$ the energies in sex-fair
matchings are $X = Y = \sqrt{N}$, in agreement with our previous
considerations.

\begin{figure}
\centerline{\epsfig{file=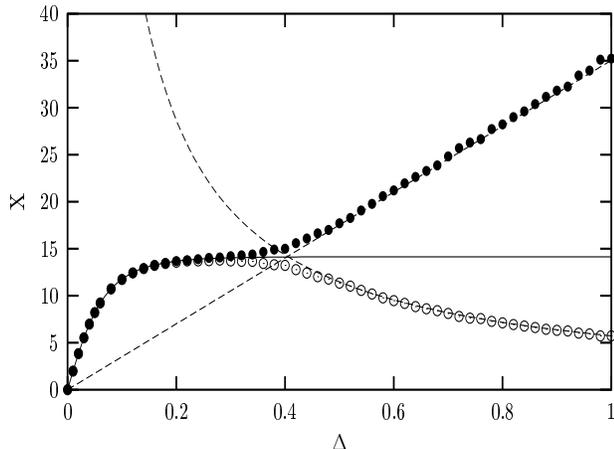,width=8cm,height=6cm,angle=0}}
\caption{Energy of men as function of $\Delta$.  
Open circles: solution of lowest energy (man-optimal GS),
filled circles: solution of highest energy (woman-optimal GS).
The solid and the dashed curves 
are the analytical results of Eq. (\ref{XYDEL}) and
Eq. (\ref{XSC}), valid for $\Delta > \Delta_{\rm c}$ 
and $\Delta < \Delta_{\rm c}$, respectively.}
\label{eofdelta}
\end{figure}

In Fig. \ref{eofdelta} we show the energy of men and women in the man-optimal 
stable matching as function of $\Delta$ for $N = 200$. Each data point
represents the average over 1000 executions of the GS algorithm.
The agreement between numerical and analytical results is very good
with the exception of the cross-over region around $\Delta_c$ where neither
of the conditions that were made to treat the problem (absence of singles
in the case $\Delta > \Delta_{\rm c}$, existence of only one stable 
matching in the case $\Delta < \Delta_{\rm c}$) is strictly fulfilled.

In the following we investigate how the number of singles depends on the
acceptance threshold $\Delta$. The probability that a woman who
receives $r$ proposals rejects all of them and remains single 
is $p_s = (1 - \Delta)^r$. Taking into account that the number of
proposals addressed to a particular woman is distributed according
to a binomial distribution
the average number of singles $N_s$ is related to the energy $X$ via
\begin{equation}
N_s(\Delta) = N \, \exp(-X \Delta)
\label{ns}
\end{equation}
(see Appendix C)  
and can be calculated explicitly from the solution of Eq.(\ref{XSC}).
The limiting cases are
\begin{equation}
N_s(\Delta) \simeq \left\{ 
\begin{array}{lll}
N \, \exp(-\Delta^2 N)  & {\rm ~~~for~} & \Delta \ll \frac{1}{\sqrt{N}} \\
N \, \exp(-\Delta \sqrt{N})  & {\rm ~~~for~} & \Delta \gg \frac{1}{\sqrt{N}}
\end{array}
\right.
\end{equation}
When $\Delta$ is decreased, the first single appears when $N_s(\Delta) = 1$.
This is the case for $\Delta = \log N / \sqrt{N} \simeq
\Delta_{\rm c}$ in accordance with our assumption that there are
no singles for $\Delta > \Delta_c$.

Finally, we generalize our results concerning the average number of stable 
matchings to the case with threshold $\Delta < 1$.
The only difference compared to the calculation for $\Delta = 1$ is that
the probability distribution of $X$ (see Appendix A) has to be replaced by
\begin{equation}
\rho(X) \simeq \frac{X^{N-1}}{\Gamma(N)} 
\left(1 - \exp(-\frac{N\Delta}{X})\right)^N
\label{rhodelta}
\end{equation}
which takes into account that all energies $x_i$ have to be smaller than $\Delta$.
Making the same approximations as before we obtain for the average number of
stable matchings
\begin{equation}
S(\Delta) \simeq \frac{N}{\rm e} \left( \log N - 2 \log(\log N)
+ 2 \log \Delta \right)
\label{NSTABDELTA}
\end{equation}
$S(\Delta)$
goes to zero at $\Delta = \log N / \sqrt{N} \simeq \Delta_{\rm c}$
in agreement with the fact that man and woman-optimal stable matchings
coincide for $\Delta < \Delta_{\rm c}$.

\section{Unequal number of men and women}

Let us now consider the more general case where the number of men and women is
not the same. To be specific, we assume that there are $N+1$ men and
$N$ women. It is obvious that in this case at least one man has to
remain single. As before, the GS algorithm can be used to 
find the man-optimal and the woman-optimal stable matching, respectively, depending
on who proposes and who judges. There is now a fundamental
asymmetry between the man and the woman-oriented version: 
when women propose the algorithm stops as usual when the last woman
has found a husband; in the man-oriented version, however, the
series of proposals stops only when the first man has reached the
bottom of his list, i.e. when he has been rejected by {\em all} women 
and must stay single.
Therefore, the members of the minority group (the women in our case) 
are substantially favored.

\begin{figure}
\centerline{\epsfig{file=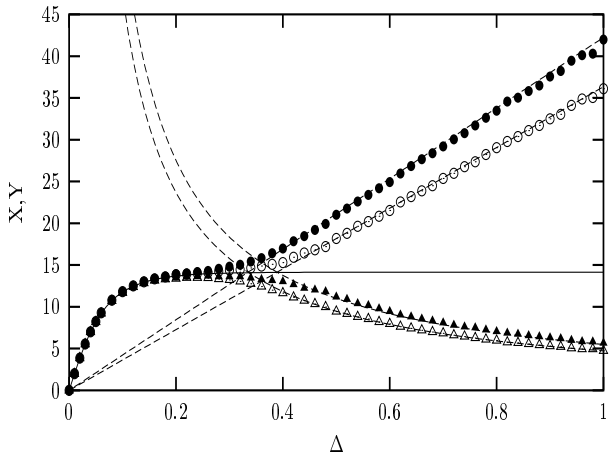,width=8cm,height=6cm,angle=0}}
\caption{Energy of men (circles) and women (triangles) as function of
$\Delta$ for 200 women and 201 men. 
Full and open symbols denote the best and the worst possible solution, respectively.
The solid and the dashed curves are the analytical results discussed in the text.}
\label{uneq}
\end{figure}

This is illustrated in Fig. \ref{uneq} where the energy of men 
(circles) and women (triangles) is
plotted versus the acceptance threshold $\Delta$. The data points 
represent averages of 1000 samples using the GS algorithm for
201 men and 200 women. Best (open symbols) and worst solutions (filled symbols)
are much closer in energy now than in the symmetric case with equal number of men and
women (see Fig. \ref{eofdelta} for comparison) indicating that the total number of stable
matchings is reduced.
The energies in all other stable matchings lie in the two narrow corridors between the  
best and the worst solution. The dashed curves 
are $X(\Delta) = X(1) \Delta$ and $Y(\Delta) = Y(1) / \Delta$
obtained from the generalization of Eq. (\ref{XYDEL})
to the present case while
the solid curve is the solution of Eq. (\ref{XSC}) valid for $\Delta < \Delta_{\rm c}$.

\section{Conclusions}
We have investigated the statistical properties of stable marriages
using both numerical and analytical tools. A detailed analysis
of the GS algorithm is provided. This algorithm yields two extreme stable matchings
and is thus particularly useful to characterize the whole set of stable
matchings.
The average number of stable matchings is calculated from an integral
representation that can be viewed as the partition sum of two species of particles
living on the unit interval,
with multiplicative interaction of their center of mass coordinates. 
Our analytical result agrees very well with numerical simulations,
in contrast to a previously obtained, less accurate formula.
In addition, the fundamental relation $X Y = N$ between the energy 
of men and women in stable matchings
follows quite naturally from our approach.
Numerical simulations 
indicate that the probability distribution of the logarithm of the
number of stable matchings is Gaussian with a universal width $\sigma_{\infty}
\approx 0.48$ for $N \rightarrow \infty$. 

When an acceptance threshold $\Delta$ is introduced,
the number of stable matchings shrinks and singles occur.
Three different regimes can be identified:
For $\Delta > \Delta_{\rm c} \simeq \log N / \sqrt{N}$
there are many stable matchings whose energies extend over a wide range and
no singles occur.
For $1/\sqrt{N} < \Delta < \Delta_{\rm c}$ there are very few or only one stable
matching. In this regime the energy of men and
women is nearly independent of $\Delta$  although the number of singles 
increases exponentially with decreasing $\Delta$.
For $\Delta < 1/\sqrt{N}$ there are many singles and only a small fraction
of married individuals.
When the number of men and women is not the same, the number of
stable matchings is also substantially reduced. Furthermore, the members
of the minority group are strongly favored compared to the majority group,
even if there is only one additional member.

The general ideas discussed in this paper could become
important for investigations on the labour market since
nowadays information about forthcoming job opportunities,
on the one hand, and qualifications of applicants, on the other hand,
can be made available to everybody via the internet.
Preliminary results on applications of the stable marriage
problem in the context of the labour market can be found in \cite{Deb99}. 
Further work on this subject is in progress. 

\acknowledgements

We would like to thank Alain Debecker for many interesting
and inspiring discussions.

\appendix

\section{Calculation of $\rho(X)$}

In this appendix we calculate the distribution function $\rho(X)$
of the variable $X = \sum_i x_i$ which is formally
given by
\begin{equation}
\rho(X) = \int_0^1 {\rm d}^Nx  \;\delta(X - \sum_i x_i)
\label{A1}
\end{equation}
According to the central limit theorem 
$\rho(X)$ converges to a normal distribution
\begin{equation}
\rho(X) = \frac{1}{\sqrt{2\pi \sigma^2}} \;
\exp(-\frac{(X- \bar X)^2}{2 \sigma^2})
\label{A2}
\end{equation}
with $\bar X = N/2$ and $\sigma^2 = N/12$ for large $N$.
This is however only true for the central part of the distribution
whereas the tails of $\rho(X)$ look very
different. Let us introduce new variables $q_i$ such that
$x_i = q_i - q_{i-1}$ for $i = 1,N$ and $q_0 = 0, q_N = X$.
Disregarding the constraint $x_i < 1$ for the moment we obtain
\begin{equation}
\rho(X) = \frac{1}{\Gamma(N)} \int_0^X {\rm d}q_1 \ldots \int_0^X {\rm d}q_{N-1}
= \frac{X^{N-1}}{\Gamma(N)}
\label{A3}
\end{equation}
where the factor $\Gamma(N)$ occurs due to the fact that the $q_i$ have to be
arranged in ascending order to satisfy the condition $x_i \ge 0$. 
The condition $x_i < 1$ can now be to implemented as follows:
Considering the $q_i$ as independent random
variables uniformly distributed
in the interval $[0,X]$ we obtain a Poisson distribution
$p(x) = \lambda \exp{(-\lambda x)}$
for the distance $x$ between two consecutive numbers, where $\lambda = N/X$.
The probability that all $x_i < 1$ is therefore given by 
\begin{equation}
P({\rm all}~ x_i<1) = \left(\int_0^1 {\rm d}x \; p(x)\right)^N = 
\left(1 - \exp{(-\frac{N}{X})}\right)^N
\label{A4}
\end{equation} 
neglecting the weak correlations among the $x_i$ due to the fact that their
sum is fixed. Combining Eqs. (\ref{A3}) and (\ref{A4}) we obtain the final result
\begin{equation}
\rho(X) = \frac{X^{N-1}}{\Gamma(N)} \left(1 - \exp{(-\frac{N}{X})}\right)^N
\end{equation}
valid for $X < N / \log N$.

\section{Average energy of women}

Let us assume that men propose in the GS algorithm and that
a woman receives $r$ proposals out of which she keeps only the best one.
If none of the offers is better than $\Delta$ she remains single at the
energy $\Delta$. The average energy thus obtained is
\begin{equation}
y  = \frac{1}{r+1} \left(1 - (1-\Delta)^{r+1}\right)
\label{B1}
\end{equation}
We have to take into account that the number $r$ of proposals that a woman receives
is not fixed but distributed according to the binomial distribution
\begin{equation}
b(r) = {R \choose r} \,
p^r (1 - p)^{R-r}
\label{B2}
\end{equation}
where $R$ is the total number of proposals that men make during the execution
of the GS algorithm and $p= 1/N$ is the probability that such an offer is
addressed to a given woman. 
Averaging Eq. (\ref{B1}) with respect to $b(r)$ yields
\begin{eqnarray}
\nonumber
\left\langle \frac{1}{r+1}\right\rangle 
& = & \sum_{r=0}^{R} {R \choose r} \, p^r (1-p)^{R-r} \frac{1}{r+1} \\
&  = &  \frac{N}{R+1} \left(1 - (1-\frac{1}{N})^{R+1}\right)
\label{B3}
\end{eqnarray}
and similarly
\begin{equation}
\left\langle \frac{(1-\Delta)^{r+1}}{r+1}\right\rangle 
  =  \frac{N}{R+1} \left((1-\frac{\Delta}{N})^{R+1} - (1-\frac{1}{N})^{R+1}\right)
\label{B4}
\end{equation}
Combining Eqs. (\ref{B1}), (\ref{B3}) and (\ref{B4}) we obtain
\begin{equation}
\langle y \rangle = \frac{N}{R+1}\left(1 - (1-\frac{\Delta}{N})^{R+1}\right)
\approx \frac{N}{R} \left(1 - \exp{(-\frac{R}{N}\Delta)} \right)
\end{equation}
The average energy of women $Y = N \langle y \rangle$ is therefore related
to the energy of men $X = R / N$ via
\begin{equation}
Y = \frac{N}{X} \left(1 - \exp{(-X\Delta)} \right)
\label{B5}
\end{equation}

\section{Number of singles}

The probability for a woman to remain single after receiving $r$ proposals is
$p_s = (1 - \Delta)^r$. Averaging with respect to the binomial distribution (\ref{B2})
yields
\begin{equation}
\langle p_s \rangle = 
\sum_{r=0}^{R} {R \choose r} \,p^r (1-p)^{R-r} (1 - \Delta)^r = 
\left(1 - \frac{\Delta}{N}\right)^R
\label{C1}
\end{equation}
where $p = 1/N$ and $R$ is the total number of proposals made. Using $X = R / N$
we obtain
\begin{equation}
\langle p_s \rangle = \exp(- X \Delta)
\label{C2}
\end{equation}

\end{document}